\documentclass[10pt, journal, letterpaper, twocolumn]{IEEEtran}

\usepackage{times, amssymb, amsmath, amsfonts, amsthm}
\usepackage{graphicx, epstopdf, subfig}
\usepackage{algorithm, algorithmic}
\usepackage{comment, color, soul}
\usepackage{mathtools}
\usepackage{cite, hyperref}

\newcommand{\beq}{\begin{equation}}
\newcommand{\eeq}{\end{equation}}
\newcommand{\tbf}{\textbf}
\newcommand{\tit}{\textit}
\newcommand{\mbf}{\boldsymbol}

\DeclareMathOperator*{\argmax}{arg\,max}
\DeclareMathOperator{\ud}{d}

\theoremstyle{plain}

\theoremstyle{plain}

\theoremstyle{plain}

\theoremstyle{plain}

\theoremstyle{plain}

\theoremstyle{plain}

\newcommand {\Ccal}{\mathcal{C}}
\newcommand {\Ecal}{\mathcal{E}}
\newcommand {\Ical}{\mathcal{I}}
\newcommand {\Kcal}{\mathcal{K}}

\newcommand {\Ncal}{\mathcal{N}}

\begin{document}

\title{Optimal Non-Coherent Detector for Ambient Backscatter Communication System}
\author{
	\IEEEauthorblockN{ 
		Sudarshan Guruacharya, \IEEEmembership{Member, IEEE}, 
		Xiao Lu, \IEEEmembership{Member, IEEE}, and
		Ekram Hossain, \IEEEmembership{Fellow, IEEE}
		}
\thanks{
	S. Guruacharya is with the Department of Electrical and Computer Engineering, New York Institute of Technology, Old Westbury, NY, USA (e-mail: sguruach@nyit.edu).
	}
\thanks{
	X. Lu is with the Department of Electrical and Computer Engineering, University of Alberta, AB, Canada (e-mail: lu9@ualberta.ca).
	}
\thanks{
	E. Hossain is with the Department of Electrical and Computer Engineering, University of Manitoba, MB, Canada (e-mail: ekram.hossain@umanitoba.ca). 
	
	This work was supported in part by a Discovery Grant from the Natural Sciences and Engineering Research Council of Canada (NSERC).
	}
}

\maketitle

\begin{abstract}
The joint probability density function (pdf) of the received signal of an ambient backscatter communication system is derived, assuming that on-off keying (OOK) is performed at the tag to form non-return to zero (NRZ) line codes, and that the ambient radio frequency (RF) signal is white Gaussian. The pdf of the received signal is then utilized to design two different types of non-coherent detectors. The first detector directly uses the received signal to perform a hypothesis test. The second detector first estimates the channel based on the observed signal and then performs the hypothesis test. Test statistics and the optimal decision threshold of the detectors are derived. The energy detector is shown to be an approximation of the second detector. For cases where the reader is able to avoid or cancel the direct interference from the RF source (e.g., through successive interference cancellation), a third detector is given as a special case of the first detector. Numerical results show that the first detector outperforms the second detector, although the second detector is computationally simpler.
\end{abstract}


\section{Introduction}
\begin{figure}[t]
	\centering
	\includegraphics[width=0.7\columnwidth]{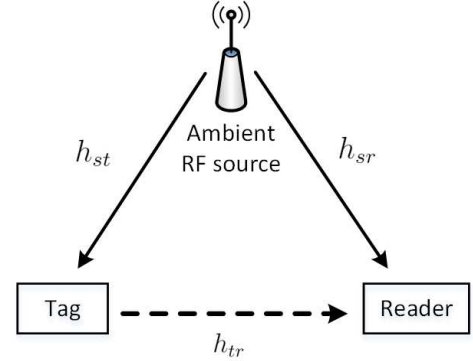}
	\caption{Ambient backscatter system consisting of RF signal source, a passive tag, and a reader.}
	\label{fig:sys-model}
\end{figure}

Ambient backscatter communication has been introduced as an energy-efficient alternative to  low-power communication systems~\cite{Griffin2009,Liu2013,H.2018Nguyen}. As shown in Fig. \ref{fig:sys-model}, in this system, a tag communicates with a reader by modulating and reflecting the ambient radio frequency (RF) signals from surrounding RF sources, such as TV stations, cellular and WiFi networks. This eliminates active RF components at the tag, leading to simpler circuitry and lower power consumption for data transmission~\cite{Liu2019,X.2018Lu} and relaying~\cite{X.Dec.2019Lu,G.Li}.

Many challenges abound in this emerging technology, one of which is the design of signal detector when the channel state information (CSI) is unknown to the reader. {The reasons for this are: 1) the wireless channel for an ambient backscatter communication system is not a traditional point-to-point channel, 2) the nature of RF signal (such as bandwidth, transmit power, waveform) exploited by the system is generally unknown to the reader and it should be considered as a random signal, and 3) because of the previous reason, the reader lacks the training symbols required to estimate the channel parameters. As such, the CSI is generally unknown to the reader.} 

A number of recent works \cite{Wang2016,Yang2016,Liu2017,Qian2017a,Qian2017b,Tao2018} have tried to address this problem. The common feature of all these works is that the detector is designed in ignorance of the statistics of the received signal when the channel is unknown. When the statistics of the received signal are used, the instantaneous CSI is assumed to be known. This leads to a semi-coherent detector, whose detection threshold depends on the randomly varying channel state. Thus, the detection threshold needs to be estimated every time the channel state changes. 
Various differential coding schemes have been used in an attempt to bypass this essential ignorance. However, references \cite{Wang2016,Yang2016,Liu2017,Qian2017a,Qian2017b} are \tit{not correct} in their claim to have built a truly non-coherent detector. To the best of our knowledge, reference \cite{Tao2018} is the only paper that has succeeded in presenting a truly non-coherent detector by using differential Manchester coding, where the CSI is not required at the reader. However, with Manchester coding, the data rate is halved. When the RF source employs orthogonal frequency-division multiplexing (OFDM), references \cite{Yang2018,ElMossallamy2019,Darsena2019} have exploited the structure of OFDM waveform to construct the required detector.

Recently, the authors in \cite{Guruacharya2019} derived the statistics for the sum of a circularly symmetric complex Gaussian (CSCG) vector with the product of a CSCG scalar and another CSCG vector. In this correspondence, we apply these results to derive the unconditional joint probability density function (pdf) of the received signal at the reader of an ambient backscatter communication system, which is then utilized to construct the \tit{correct} optimal non-coherent detector, which does not require instantaneous CSI,  when non-return-to-zero (NRZ) coding is used.\footnote{Note that the result is valid for return-to-zero (RZ) line code as well.} Three different detectors are presented, whose detection thresholds are constants, and their performances are studied.

{\em Notations:}  Lower/upper boldface letters denote vectors/matrices;  $|\cdot|$, $\tbf{I}_N$, $||{\cdot}||$ denotes magnitude, identity matrix of size $N$, and Euclidean norm; $p(\cdot)$ and $\Pr(\cdot)$ denotes pdf and probability operator, respectively. $\Ccal\Ncal(0,\sigma^2 \tbf{I}_N)$ denotes a CSCG distribution with zero-mean and co-variance matrix $\sigma^2 \tbf {I}_N$; $\gamma(s,x)$ and $\Gamma(s,x)$ denotes lower and upper incomplete gamma functions, respectively.

\section{System Model}
\label{sec:sys-model}

Consider a simple ambient backscatter configuration which consists of an ambient RF source, a passive tag, and a reader as shown in Fig. \ref{fig:sys-model}. The RF energy broadcasted by the source is received by both the tag and the receiver. The passive tag can reflect the incoming RF signal to the reader by changing its impedance. As such, the tag is capable of transmitting binary symbols to the reader by choosing whether or not to backscatter the incident RF energy. The symbols ``0'' and ``1'' correspond to the tag's non-backscattering and backscattering state. The reader senses the changes to its received signal and decodes the transmitted symbols of the tag.

The baseband signal received at the tag at the $n$-th sampling instance is 
\vspace*{-0.3em}
\beq
x[n] = h_{st} s[n],
\eeq
where $s[n]$ is the unknown random RF signal and $h_{st}$ represents the channel coefficient between the RF source and the tag. Since the thermal noise at the tag is very small, we will follow the convention where this noise is omitted. We assume that $s[n]$ is a complex white Gaussian signal. That is, the signals are independent and identically distributed (i.i.d.) as $\tbf{s} \sim \Ccal\Ncal(0,\sigma_s^2 \tbf{I}_N)$ where $\tbf{s} = (s[0], \ldots, s[N-1])^T$. Commonly used modulation schemes, such as OFDM and code-division multiplexing (CDM), are approximately white Gaussian in time domain \cite{Wei2010}. Due to the central limit theorem, this is also a good approximation when the exploited RF signal results from the superposition of signals from multiple RF sources. Likewise, we assume that we have a scatter-rich environment, such as indoor home or office spaces, allowing us to model $h_{st}$ as Rayleigh fading channel whose distribution is given by $h_{st} \sim \Ccal\Ncal(0,\sigma_{st}^2)$. 

Let us denote the $k$-th binary symbol of the tag as $b_k \in \{0,1\}$, which is assumed to be equiprobable. The tag transmits data at a slower rate than the RF signal. As such, we can assume that $b_k$ is a constant over the interval of observation where $N$ samples are collected. Assuming that the tag uses NRZ line code to represent the bits via simple on-off keying (OOK), the signal backscattered by the tag is given by
\vspace*{-0.3em}
\beq
x_b[n] = \alpha b_k x[n],
\eeq  
where $\alpha \in (0,1)$ is a scaling term related to the scattering efficiency and antenna gain of the tag.  
We assume that the data bits are framed by start and stop bits, and that the transmission is asynchronous.

The baseband signal received at the reader corresponding to the $k$-th tag symbol $b_k$ is 
\begin{align}
y[n] &= h_{sr} s[n] + h_{tr} x_b[n] + w[n], \nonumber \\
&= (h_{sr} + \alpha b_k h_{st} h_{tr}) s[n] + w[n].
\end{align}
Here $h_{sr}$ is the channel coefficient between the RF source to the reader, while $h_{tr}$ is the channel coefficient between the tag to the reader. We will assume that $N$ samples of $y$ fall within a single OOK symbol duration. We will also assume both $h_{sr}$ and $h_{tr}$ to be Rayleigh fading channels; thus $h_{sr} \sim \Ccal\Ncal(0, \sigma_{sr}^2)$ and $h_{tr} \sim \Ccal\Ncal(0, \sigma_{tr}^2)$. Likewise, $\tbf{w} \sim \Ccal\Ncal(0, \sigma_w^2 \tbf{I}_N)$ is the i.i.d. additive complex white Gaussian noise, where $\tbf{w} = (w[0], \ldots, w[N-1])^T$. Hence, depending on the value of $b_k$, signal received at the reader is 
\begin{align}
y[n] = \left\{ \begin{array}{ll} 
				 h_0  s[n] + w[n]  & \mathrm{if} \;\; b_k =0 \\
				 h_1 s[n] + w[n] &  \mathrm{if} \;\; b_k = 1
			   \end{array},
 		\right.
\end{align}
where $h_0 = h_{sr}$ and $h_1 = h_{sr} + \alpha h_{st} h_{tr}$.

Let $\tbf{y} = (y[0], \ldots, y[N-1])^T$ be a vector of $N$ observations sampled at the reader. In order to construct an  optimal non-coherent detector, we need to know what the distribution of $\tbf{y}$ is. However, the answer to this question depends on the coherence time of the wireless channels. In the simplest instance, we will assume that the channel coherence time is equal to the observation time so that the channel coefficients $h_{st}$, $h_{tr}$, and $h_{sr}$ remain constant during the $N$ observations, but may vary in different coherence intervals independently. Another complication is that since any two samples $y[m]$ and $y[n]$ share the same random channel values for any $m \neq n$, this implies that $y[m]$ and $y[n]$ are no longer independent of each other.

\section{Joint PDF of $\mbf{y}$}
\label{sec:analysis}
In this section, we will derive the joint pdf of $\tbf{y}$ unconditioned from $h_0$ and $h_1$, which is required during the design of a non-coherent detector. We will use the integral function $\Ical_L(z; a, b)$ and its properties, reviewed in the \tbf{Appendix}, to express the pdf of $\tbf{y}$ in a convenient form.  Let the squared magnitude of channel coefficients be denoted as $v_0 = |h_0|^2$ and $v_1 = |h_1|^2$. 

When $b_k = 0$, the conditional pdf of $y[n]$ given $v_0$ is 
\beq
y[n] \, \big| \, v_0 \sim \Ccal\Ncal(0, v_0 \sigma_s^2 + \sigma_w^2),
\eeq
where the pdf of $v_0$ is given by the exponential distribution
\beq
p_{V_0}(v_0) = \frac{1}{\sigma_{sr}^2} e^{-v_0/\sigma_{sr}^2}.
\label{eqn:pdf-v0}
\eeq
De-conditioning with respect to $v_0$, the pdf of $\tbf{y}$ can be directly written as \cite[Eqn 4]{Guruacharya2019}
\beq
p_{\tbf{Y}}(\tbf{y} | b_k =0) = K_0 \Ical_N(\|\tbf{y}\|^2; \sigma_w^2, \sigma_{sr}^2 \sigma_s^2),
\label{eqn:pdf-y-b=0}
\eeq
where $K_0 = e^{\frac{\sigma_w^2}{\sigma_{sr}^2 \sigma_s^2} } \big/ (\pi^N \sigma_{sr}^2 \sigma_s^2)$ is a constant.

When $b_k=1$, the pdf of $y[n]$ conditioned on $v_1$ is 
\beq
y[n] \, \big| \, v_1 \sim \Ccal\Ncal(0, v_1 \sigma_s^2 + \sigma_w^2).
\eeq
Hence, the conditional pdf of vector $\tbf{y}$ is
\beq
p_{\tbf{Y}}(\tbf{y}| v_1) = \frac{1}{(\pi (v_1 \sigma_s^2 + \sigma_w^2))^N} \exp\left( - \frac{\|\tbf{y}\|^2}{v_1 \sigma_s^2 + \sigma_w^2} \right).
\label{eqn:pdf-y-given-h}
\eeq
Using \cite[Eqn 15]{Guruacharya2019}, the pdf of $v_1$ is
\beq
p_{V_1}(v_1) = \frac{1}{\alpha^2 \sigma_{st}^2 \sigma_{tr}^2 } \exp\left( \frac{\sigma_{sr}^2}{ \alpha^2 \sigma_{st}^2 \sigma_{tr}^2} \right)  \Ical_1(v_1; \sigma_{sr}^2, \alpha^2 \sigma_{st}^2 \sigma_{tr}^2).
\label{eqn:pdf-v1}
\eeq 
De-conditioning (\ref{eqn:pdf-y-given-h}), we have
\beq
p_{\tbf{Y}}(\tbf{y}| b_k =1) = \int_0^\infty p_{\tbf{Y}}(\tbf{y}| v_1 ) p_{V_1}(v_1) \ud v_1.
\label{eqn:pdf-y-int}
\eeq
Let $u = v_1 \sigma_s^2 + \sigma_w^2$, then substituting (\ref{eqn:pdf-y-given-h}) and  (\ref{eqn:pdf-v1}) into (\ref{eqn:pdf-y-int}), we obtain 
\begin{align}
p_{\tbf{Y}}(\tbf{y}|b_k=1) &= K_1 \int_{\sigma_w^2}^\infty \frac{e^{-\|\tbf{y}\|^2/u}}{u^N}  \nonumber \\ 
& \qquad \times \Ical_1\left( \frac{u-\sigma_w^2}{\sigma_s^2}; \sigma_{sr}^2, \alpha^2 \sigma_{st}^2 \sigma_{tr}^2 \right) \ud u, 
\label{eqn:pdf-y-b=1}
\end{align}
where $K_1 = e^{\frac{\sigma_{sr}^2}{ \alpha^2 \sigma_{st}^2 \sigma_{tr}^2}} \big/ (\pi^N \alpha^2 \sigma_{st}^2 \sigma_{tr}^2 \sigma_s^2)$ is a constant.

\section{Optimal non-coherent detector}
\label{sec:detector}
Here we will give two approaches to construct a detector for $b_k$. The first approach directly utilizes the pdf of $\tbf{y}$. The second approach estimates $v$ and then performs the hypothesis test, given the estimate of $v$. A third detector is also given as a special case of the first detector, assuming that the direct link interference has somehow been nullified. In this section, we will denote the energy of signal $\tbf{y}$ by $z = \|\tbf{y}\|^2$.

\subsection{First Detector: Direct Approach}
\label{subsec:first-detector}
The simplest approach in constructing a detector is to directly use the pdf of $\tbf{y}$ to derive the test statistics and decision threshold. Thereby, given the observation vector $\tbf{y}$, the detector needs to perform a binary hypothesis test where $H_0 : \tbf{y} = h_0 \tbf{s} + \tbf{w}$ and $H_1 : \tbf{y} = h_1 \tbf{s} + \tbf{w}$. The optimal non-coherent detector is given by the likelihood ratio test (LRT) detector. We have the LRT given by 
\beq
\frac{p_{\tbf{Y}}(\tbf{y} | b_k =1) }{p_{\tbf{Y}}(\tbf{y} | b_k =0)} \overset{H_1}{\underset{H_0}{\gtrless}} 1.
\label{def:LLR-1}
\eeq

From (\ref{eqn:pdf-y-b=0}) and (\ref{eqn:pdf-y-b=1}),  the test statistics for log-LRT is
\begin{align}
\Lambda_1 &= \log \int_{\sigma_w^2}^\infty \frac{e^{-z/u}}{u^N} \Ical_1\left( \frac{u-\sigma_w^2}{\sigma_s^2}; \sigma_{sr}^2,  \alpha^2 \sigma_{st}^2 \sigma_{tr}^2 \right) \ud u  \nonumber \\
& \qquad \qquad - \log \Ical_N(z; \sigma_w^2, \sigma_{sr}^2 \sigma_s^2),
\label{eqn:test-stat-1}
\end{align}
with the optimal decision threshold as $\theta_1^* = \log(\frac{K_0}{K_1})$, or
\beq 
\theta_1^* = \frac{\sigma_w^2}{\sigma_{sr}^2 \sigma_s^2} - \frac{\sigma_{sr}^2}{\alpha^2 \sigma_{st}^2 \sigma_{tr}^2} + \log\left( \frac{\alpha^2 \sigma_{st}^2 \sigma_{tr}^2}{\sigma_{sr}^2} \right). 
\label{eqn:opt-thres-1}
\eeq

Let $\widehat{b}_k$ be the decision made by the detector, then the detector will decide $\widehat{b}_k = 0$ if $\Lambda_1 < \theta_1^*$, otherwise $\widehat{b}_k = 1$ if $\Lambda_1 > \theta_1^*$.

\subsection{Second Detector: Indirect Approach}
\label{subsec:second-detector}
Another approach to construct a non-coherent detector is as follows: let $h = h_{sr} + b_k h_{st} h_{tr}$ be the unknown channel that depends on $b_k$ and let $v = |h|^2$. For a given $v$, we know that 
\[ p_{\tbf{Y}}(\tbf{y} | v) = \frac{1}{(\pi(v \sigma_s^2 + \sigma_w^2))^N} \exp \left(-\frac{z}{v \sigma_s^2 + \sigma_w^2} \right). \]
The maximum log-likelihood estimate of $v$ is 
\beq
v^* = \argmax_{v \geq 0} \; \log p_{\tbf{Y}}(\tbf{y} | v),  
\label{eqn:ml-v}
\eeq
which after some basic calculus is given by 
\beq
v^* = \left(\frac{z}{N \sigma_s^2} - \frac{\sigma_w^2}{\sigma_s^2} \right)_+,
\label{eqn:v-estimate}
\eeq
where $(x)_+ = \max(0, x)$. Using the estimate $v^*$ we can decide whether $H_0 : v^* = |h_0|^2$ or $H_1 : v^* = |h_1|^2$ is true. The LRT detector is given by 
\beq
\frac{p_{V}(v^* | b_k =1) }{p_{V}(v^* | b_k =0)} \overset{H_1}{\underset{H_0}{\gtrless}} 1.
\label{def:LLR-2}
\eeq
Here $p_{V}(v | b_k =0)$ is given by (\ref{eqn:pdf-v0}) and $p_{V}(v | b_k=1)$ is given by (\ref{eqn:pdf-v1}). Hence, the log-LRT will give us the test statistics
\beq
\Lambda_2 = \frac{v^*}{\sigma_{sr}^2} + \log \Ical_1 (v^*; \sigma_{sr}^2, \alpha^2 \sigma_{st}^2 \sigma_{tr}^2), 
\label{eqn:test-stat-2}
\eeq
with an optimal decision threshold 
\beq
\theta_2^* = \log \left( \frac{\alpha^2 \sigma_{st}^2 \sigma_{tr}^2}{\sigma_{sr}^2} \right) - \frac{\sigma_{sr}^2}{\alpha^2 \sigma_{st}^2 \sigma_{tr}^2}.
\label{eqn:opt-thres-2}
\eeq

As done previously, the detector will decide $\widehat{b}_k = 0$ if $\Lambda_2 < \theta_2^*$, otherwise $\widehat{b}_k = 1$ if $\Lambda_2 > \theta_2^*$.

\subsection{Discussions}
\label{subsec:discussion}
\begin{enumerate}
	\item In the indirect approach, if we neglect $\sigma_w^2$ and $\log \Ical_1(v^*; \sigma_{sr}^2, \alpha^2 \sigma_{st}^2 \sigma_{tr}^2)$ in the test statistics, the second detector will reduce to a simple \tit{energy detector} with $\Lambda_2 \approx z/N \sigma_{sr}^2 \sigma_s^2$. The decision threshold for this energy detector is given by (\ref{eqn:opt-thres-2}). Hence, the energy detector will have similar behavior and limitations as the second detector.
	\item When $\sigma_w^2/\sigma_{sr}^2 \sigma_s^2$ is negligible, then $\theta^*_1 = \theta^*_2$.
\end{enumerate}

\subsection{Special Case: Detection After Interference Nullification}
\label{subsec:special-case}
In some cases, the reader can avoid or cancel the direct interference from the RF source. For example, if the reader can decode the symbols transmitted by the RF source, then it can cancel the interference to the received backscattered signal via successive interference cancellation (SIC) technique. Likewise, the backscattered signal can be modulated to appear in a different frequency band, free from direct interference \cite{ElMossallamy2019}. Assuming that direct RF interference is somehow nullified, we can construct an optimal detector for this special case using the results obtained so far.  When $b_k =0$, there is only noise at the reader. Thus, the pdf of $\tbf{y}$ is simply
\beq
p_{\tbf{Y}}(\tbf{y} | b_k=0) = \frac{e^{-z/\sigma_w^2}}{(\pi \sigma_w^2)^N}.
\eeq
When $b_k=1$ the pdf of $\tbf{y}$ is given by (\ref{eqn:pdf-y-b=1}), where we set $\sigma_{sr}^2 = 0$, which after applying (\ref{eqn:integral-a=0}), we obtain
\begin{align}
& p_{\tbf{Y}}(\tbf{y} | b_k=1) = \frac{2}{\pi^N \alpha^2 \sigma_{st}^2 \sigma_{tr}^2 \sigma_s^2} \nonumber \\
& \hspace{20mm} \times \int_{\sigma_w^2}^\infty \frac{e^{-z/u}}{u^N} \Kcal_0 \left( 2 \sqrt{\frac{u - \sigma^2_w}{\alpha^2 \sigma_{st}^2 \sigma_{tr}^2 \sigma_s^2}} \right) \ud u,
\end{align}
where $\Kcal_0$ is the zeroth-order modified Bessel function of the second kind. The log-LRT in (\ref{def:LLR-1}) will yield the test statistics
\beq
\Lambda_3 = \frac{z}{\sigma_w^2} + \log \int_{\sigma_w^2}^\infty \frac{e^{-z/u}}{u^N} \Kcal_0 \left( 2 \sqrt{\frac{u - \sigma^2_w}{\alpha^2 \sigma_{st}^2 \sigma_{tr}^2 \sigma_s^2}} \right) \ud u,
\label{eqn:test-stat-3}
\eeq
with an optimal decision threshold 
\beq
\theta^*_3 = \log \left(\frac{\alpha^2 \sigma_{st}^2 \sigma_{tr}^2 \sigma_s^2}{2\sigma_w^2} \right) - (N-1) \log \sigma_w^2.
\eeq


\section{Numerical Evaluation}
\label{sec:numerical}

\begin{figure}[t]
	\centering
	\includegraphics[width=0.95\columnwidth]{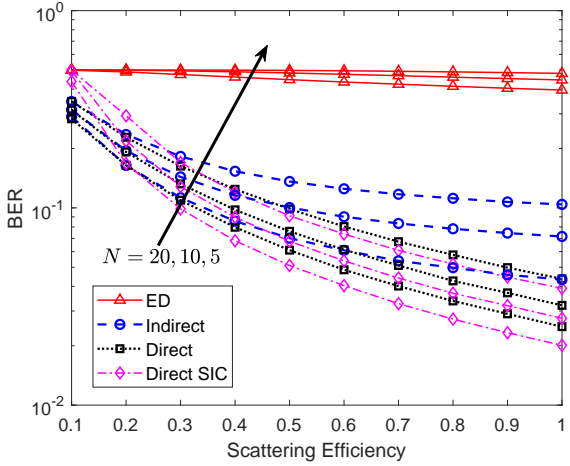}
	\caption{BER versus $\alpha$ given INR $\eta_2 = -10$ dB and $\sigma_{tr}^2 = 30$ dB. }
	\label{fig:BER_SC}
\end{figure}

\begin{figure}[t]
	\centering
	\includegraphics[width=0.95\columnwidth]{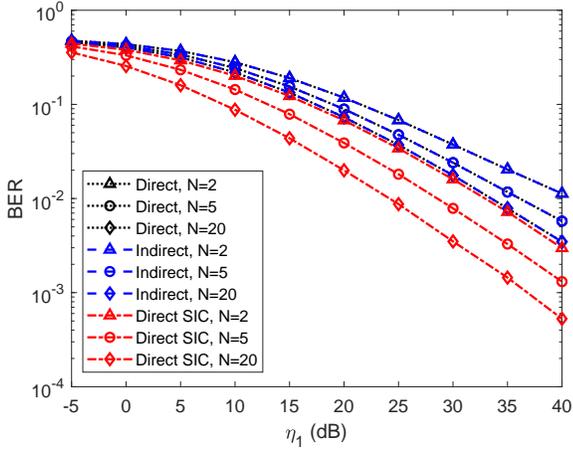}
	\caption{BER versus SIR ($\eta_1$) given INR $\eta_2 = 0$ dB. }
	\label{fig:ber-vs-sinr}
\end{figure}

\begin{figure}[t]
	\centering
	\includegraphics[width=0.95\columnwidth]{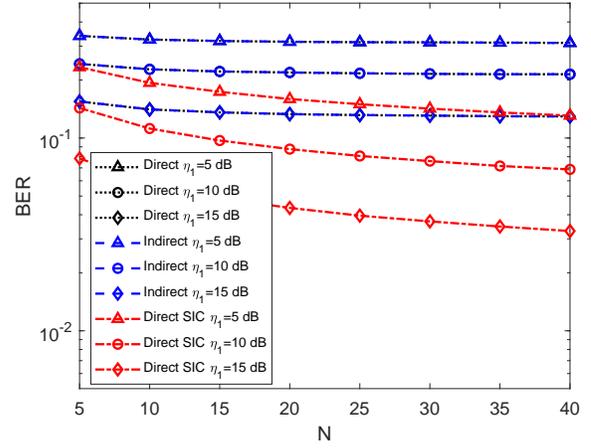}
	\caption{BER versus $N$ given INR $\eta_2 = 0$ dB. }
	\label{fig:ber-vs-N}
\end{figure}

\begin{figure}[t]
	\centering
	\includegraphics[width=0.95\columnwidth]{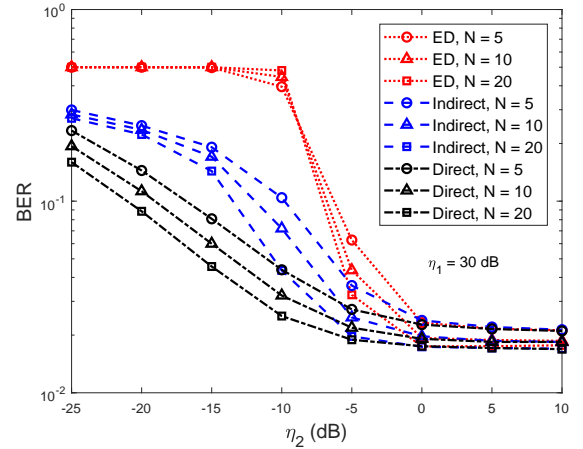}
	\caption{BER versus INR $(\eta_2)$ given SIR $\eta_1 = 30$ dB. }
	\label{fig:ber-vs-inr}
\end{figure}

Let the decision made by the detector for the $k$-th transmitted bit be denoted as $\widehat{b}_k$, then the bit error rate (BER) of the detector is defined as 
\beq
\mathrm{BER} = \frac{1}{2}\left[ \Pr(\widehat{b}_k = 0 | b_k =1) + \Pr(\widehat{b}_k = 1 | b_k =0) \right].
\eeq
Since the signal from the direct source-to-reader path appears as an interference to the desired signal from the source-tag-reader path, in the following, we will plot the BER of the non-coherent detector with respect to the signal-to-interference-plus-noise ratio (SINR). The SINR is defined as 
\beq
\eta = \frac{ \alpha^2 \sigma_{st}^2 \sigma_{tr}^2 \sigma_s^2}{\sigma_{sr}^2 \sigma_s^2 + \sigma_w^2} = \frac{\eta_1}{1 + \frac{1}{\eta_2}},
\eeq
where $\eta_1 = \alpha^2 \sigma_{st}^2 \sigma_{tr}^2/\sigma_{sr}^2$ is the signal-to-interference ratio (SIR) while $\eta_2 = \sigma_{sr}^2 \sigma_s^2 / \sigma_w^2$ is the interference-to-noise ratio (INR). 
Given these definitions, we can re-express the optimal thresholds for the detectors as $\theta_1^* = 1/\eta_2  - 1/\eta_1 + \log(\eta_1)$, $\theta_2^* = \log(\eta_1) - 1/\eta_1$, and $\theta^*_3 = \log (\eta_1 \eta_2/2) - (N-1) \log \sigma_w^2$ which are constants. Here, the $\eta$, $\eta_1$ and $\eta_2$ are readily estimated without knowing the channel variances.

In practice, the tag and the reader are separated by a short distance while the RF source is located far away from both the tag and the reader. As such, we can approximately consider the tag and the reader to be equidistant from the RF source. Hence, $\sigma_{st}^2 = \sigma_{sr}^2$ and $\eta_1 = \alpha^2 \sigma_{tr}^2$. In the Monte Carlo simulations, we change $\eta_1$ by varying either $\alpha$ or  $\sigma_{tr}^2$. Each point in a plot is constructed using $10^6$ Monte Carlo instances. To speed up the simulation, we evaluate integrals in (\ref{eqn:test-stat-1}), (\ref{eqn:test-stat-2}) and (\ref{eqn:test-stat-3}) by first creating a lookup table for $z = 0, \Delta, 2 \Delta, \ldots, z_{\max}$, where $\Delta$ is the table's step size. The lookup table is then used to interpolate the integral value for required $z \in [0, z_{\max}]$ during the Monte Carlo simulation. We set $\Delta = 0.1$ and $z_{\max} = 2000$ and use linear interpolation. If $z > z_{\max}$, we use (\ref{ineq:bessel-upper-bound}) as an approximation for $\Ical_L(z)$ in (\ref{eqn:test-stat-2}), or numerically evaluate the integrals. 

{
Fig. \ref{fig:BER_SC} depicts the impact of scattering efficiency $\alpha$ on different detection schemes in terms of BER. Here we have set $\sigma_{tr}^2 = 30$ dB and $\eta_2 = -10$ dB. We observe that BER monotonously decreases with $\alpha$ for all three detection schemes. When $\alpha$ is small, the three proposed detectors render comparable performance. However, as $\alpha$ increases, the first detector (labeled as ``Direct") results in considerable lower BER than the second detector (``Indirect").
Moreover, the performance gap between the first detector and the third detector (labeled as ``Direct SIC") keeps increasing with $\alpha$. For the sake of comparison, we have plotted the performance of an energy detector (labeled as ``ED"), see Sec. \ref{subsec:discussion}. We see that all three proposed detectors considerably outperforms the energy detector for larger values of $\alpha$. In the following section, we set $\alpha=1$ which gives the best performance for all the detection schemes and $\eta_2 = 0$ dB.}

In Fig. \ref{fig:ber-vs-sinr}, we plot the BER versus SIR. We find that both the first and second detectors have similar performances, while the third detector unsurprisingly outperforms the first two detectors. We observe that with increasing SIR, the BER decreases. As we vary the sample size $N$, we observe that the BER decreases with increasing $N$  for a given SIR. However, as $N$ increases, the BER saturates and does not improve beyond a certain point. The change in BER with respect to sample size $N$ is more obvious in Fig. \ref{fig:ber-vs-N}, where the contrast in the behavior of first/second and third detectors is distinctly observed. 

Next, in Fig. \ref{fig:ber-vs-inr} BER versus INR is plotted, where $\eta_1 = 30$ dB and $\eta_2$ is changed by varying $\sigma_s^2$. The first and second detectors are compared with the energy detector. For all three detectors, BER decreases with increasing INR and saturates at some level. We also observe that for large INR, all three detectors have similar performances. However, in the lower INR regime, $\eta_2 < 0$ dB, the first detector is better than the second detector, which in turn is better than the energy detector. The leveling effect on BER in Fig. \ref{fig:ber-vs-inr} can be explained by the fact that the tag exploits an ambient RF signal to transmit its message, which is also superimposed as unwanted interference at the reader. Therefore, when the RF signal power $\sigma_s^2$ increases, the SINR $\eta$ saturates to $ \lim_{\sigma_s^2 \to \infty} \eta = \frac{\alpha^2 \sigma_{st}^2 \sigma_{tr}^2}{\sigma_{sr}^2} = \eta_1$, which is a constant.

{These results imply that we cannot improve the performance of a non-coherent ambient backscattering system indefinitely by simply increasing the transmit power of the ambient RF source, without interference avoidance/cancellation, or by increasing the number of samples. Performance can be improved by increasing the SIR, which in the context of source-tag-reader geometry implies that tag-to-reader distance be as small as possible.}

\section{Conclusion}
\label{sec:conclusion}
We have derived the joint pdf of the received signal at the tag of an ambient backscatter communication system, which was then used to design two different types of detectors. The energy detector was shown to be an approximation of the second detector, while the interference-free case was also studied. Numerical results show that the first detector outperforms the second detector under low INR, but have similar performance under high INR. There is only a limited improvement in the performance of non-coherent detectors with an increase in sample size. However, the second detector is computationally simpler than the first detector. In practice, lookup tables are essential for fast processing. The work can be extended to different channel models, higher modulation schemes, and multiple antenna systems.

\section*{Appendix: Integral function $\Ical_L(z; a, b)$}

The integral function $\Ical_L(z; a, b)$ is defined as \cite{Guruacharya2019}
\beq
\Ical_L(z; a, b) = \int_a^\infty \frac{1}{t^L} \exp\left\{- \left( \frac{z}{t} + \frac{t}{b} \right) \right\} \ud t,
\label{def:integral}
\eeq
where $z\geq0$, $a \geq 0$, and $b\geq0$. This function (\ref{def:integral}) is closely related to incomplete Bessel function \cite{Harris2008}. While $L$ can be any real number, for our purpose, we will assume it to be a positive integer.

 Since we assume $a$ and $b$ to be constants, when it is unambiguous, we will denote $\Ical_L(z; a, b)$ by $\Ical_L(z)$. It is easy to verify that $\Ical_L(z)$ is a positive, decreasing function; and its values at zero and infinity are given by
\beq
\Ical_L(0) = \frac{1}{a^{L-1}} \Ecal_L\left(\frac{a}{b}\right) \quad\mathrm{and}\quad \Ical_L(\infty) = 0,
\label{eqn:specific-values-integral-I}
\eeq
where $\Ecal_n(x) = x^{n-1} \int_x^\infty t^{-n} e^{-t} \ud t = x^{n-1} \Gamma(1-n,x)$ is the generalized exponential integral function \cite[Ch. 8.19]{Olver2010}.  

The $\Ical_L(z)$ has the following limiting forms \cite{Guruacharya2019}: 
\begin{align}
\lim_{a \to 0} \Ical_L(z) &= \frac{2}{(bz)^{\frac{L-1}{2}}} \Kcal_{L-1} \left( 2 \sqrt{\frac{z}{b}} \right), \label{eqn:integral-a=0} \\
\lim_{b\to\infty} \Ical_L(z) &= \frac{1}{z^{L-1}} \gamma\left(L-1, \frac{z}{a} \right),
\label{eqn:integral-b=infty}
\end{align}
where $\Kcal_{\nu}$ is the $\nu$-th order modified Bessel function of the second kind. Likewise, $\lim_{a \to \infty} \Ical(z) = \lim_{b \to 0} \Ical(z) = 0$. 

The derivative of $\Ical_L(z)$ is $\Ical_L'(z) = -\Ical_{L+1}(z)$, which can be used with (\ref{eqn:specific-values-integral-I}) to obtain the Taylor expansion of $\Ical_L(z)$ at the origin as \cite{Guruacharya2019}
\vspace*{-0.5em}
\beq
\Ical_L(z) = \sum_{n=0}^\infty \frac{(-1)^n}{a^{L+n-1}} \Ecal_{L+n} \left( \frac{a}{b} \right) \frac{z^n}{n!}.
\eeq 

From (\ref{def:integral}) and (\ref{eqn:integral-a=0}), we clearly have the inequality
\beq
\Ical_L(z) \leq \frac{2}{(bz)^{\frac{L-1}{2}}} \Kcal_{L-1} \left( 2 \sqrt{\frac{z}{b}} \right).
\label{ineq:bessel-upper-bound}
\eeq

We can also derive a lower and an upper bound as follows: In the definition (\ref{def:integral}), since $e^{-z/a} \leq e^{-z/t} \leq 1$, we have the lower bound
\vspace*{-1em}
\beq
\Ical_L(z) \geq e^{-z/a} \int_a^\infty \frac{1}{t^L}e^{-t/b} \ud t = \frac{e^{-z/a}}{a^{L-1}} \Ecal_L \left( \frac{a}{b} \right).
\label{ineq:lower-bound}
\eeq
Note that (\ref{ineq:lower-bound}) is exact at $z=0$. Similarly, since $e^{-a/b} \geq e^{-t/b}$, we have an upper bound as
\vspace*{-0.2em}
\beq
\Ical_L(z) \leq e^{a/b} \int_a^\infty \frac{e^{-z/t}}{t^L} \ud t = \frac{e^{-a/b}}{z^{L-1}} \gamma\left(L-1, \frac{z}{a} \right).
\label{ineq:upper-bound}
\eeq
Note that (\ref{ineq:upper-bound}) is infinite when $L=1$. Thus, from (\ref{ineq:lower-bound}) and (\ref{ineq:upper-bound}) we have the inequality for $\Ical_L(z)$ as 
\beq
\frac{e^{-z/a}}{a^{L-1}} \Ecal_L \left( \frac{a}{b} \right) \leq \Ical_L(z) \leq \frac{e^{-a/b}}{z^{L-1}} \gamma\left(L-1, \frac{z}{a} \right).
\label{ineq:upp-low-bound}
\eeq

For small values of $z$, the relative error of the two bounds in (\ref{ineq:upp-low-bound}) are small; but the relative error increase as $z$ increases. In contrast,  (\ref{ineq:bessel-upper-bound}) is best for large values of $z$, in the sense that its relative error goes to zero as $z$ increase. These three bounds can be used as approximations to $\Ical_L(z)$. Note that these inequalities have not appeared in \cite{Guruacharya2019}.


\begin{thebibliography}{1}
	\bibitem{Griffin2009} J. D. Griffin and G. D. Durgin, ``Complete link budgets for backscatter-radio and RFID systems,'' {\em  IEEE Antennas and Propagation Mag.}, vol. 51, no. 2, pp. 11-25, Apr. 2009.

	
	\bibitem{Liu2013} V. Liu, {\em et al.}, ``Ambient backscatter: Wireless communication out of thin air,'' {\em ACM SIGCOMM Computer Commun. Review}, vol. 43, no. 4, pp. 39-50, Oct. 2013.

	\bibitem{H.2018Nguyen}
	V. H. Nguyen, {\em et al.}, ``Ambient backscatter communications: A contemporary survey," \emph{IEEE Communications Surveys $\&$ Tutorials}, vol. 20, no. 4, pp. 2889-2922, Fourthquarter 2018. 

	
	\bibitem{Liu2019} W. Liu, {\em et al.}, ``Next generation backscatter communication: Systems, techniques, and applications,'' {\em EURASIP J. Wireless Commun. and Networking}, vol. 2019, no. 69, pp. 1-12, 2019.
	
	\bibitem{X.2018Lu}
	X. Lu, \emph{et al.}, ``Ambient backscatter assisted wireless powered communications," \emph{IEEE Wireless Communications}, vol. 25, no. 2, 170-177,  Apr. 2018.

	\bibitem{X.Dec.2019Lu} 
	X. Lu, H. Jiang, D. Niyato, E. Hossain, and P. Wang, "Ambient backscatter-assisted wireless-powered relaying," \emph{IEEE Transactions on Green Communications and Networking}, vol. 3, no. 4, pp. 1087-1105, Dec. 2019.
	
	\bibitem{G.Li}
	G. Li, X. Lu, D. Niyato
	"Bandit Approach for Mode Selection in Ambient Backscatter-Assisted Wireless-Powered Relaying,"
	\emph{IEEE Transactions on Vehicular Technology}, to appear.


	\bibitem{Wang2016} G. Wang, {\em et al.}, ``Ambient backscatter communication systems: Detection and performance analysis," {\em IEEE Trans. Commun.}, vol. 64, no. 11, pp. 1-10, Aug. 2016.
	
	\bibitem{Yang2016} G. Yang and Y.-C. Liang, ``Backscatter communications over ambient OFDM signals: Transceiver design and performance analysis,'' {\em 2016 IEEE Global Communications Conference (GLOBECOM)}, pp. 1-4, 4-8 Dec., 2016.
	
	\bibitem{Liu2017} Y. Liu {\em et al.}, ``Coding and detection schemes for ambient backscatter communication systems," {\em IEEE Access}, vol. 5, no. 99, pp. 4947-4953, Mar. 2017.
	
	\bibitem{Qian2017a} J. Qian, {\em et al.}, ``Noncoherent detections for ambient backscatter system,'' {\em IEEE Trans. Wireless Commun.}, vol. 16, no. 3, pp. 1412-1422, Mar. 2017. 
	
	\bibitem{Qian2017b} J. Qian, {\em et al.}, ``Semi-coherent detection and performance analysis for ambient backscatter system,'' {\em IEEE Trans. Wireless Commun.}, vol. 65, no. 12, pp. 5266-5278, Dec. 2017. 
	
	\bibitem{Tao2018} Q. Tao, {\em et al.}, ``Symbol detection of ambient backscatter systems with Manchester coding,'' {\em IEEE Trans. Wireless Commun}, vol. 17, no. 6, pp. 4028-4038, Jun. 2018.
	
	 \bibitem{Yang2018} G. Yang, {\em et al.}, ``Modulation in the air: Backscatter communication over ambient OFDM carrier", {\em IEEE Trans. Commun.}, vol. 66, no. 3, pp. 1219-1233, Mar. 2018.
	
	\bibitem{ElMossallamy2019} M. A. ElMossallamy, {\em et al.}, ``Noncoherent backscatter communications over ambient OFDM signals," {\em IEEE Trans. Commun.}, vol. 67, no. 5, pp. 3597-3611, May 2019.
	
	\bibitem{Darsena2019} D. Darsena, ``Noncoherent detection for ambient backscatter communications over OFDM signals," {\em IEEE Access}, vol. 7, 2019.
	
	\bibitem{Guruacharya2019} S. Guruacharya, B. K. Chalise, and B. Himed, ``On the product of  complex Gaussians  with  applications  to  radar,'' {\em IEEE Signal Process. Lett.}, vol. 26, no. 10, pp. 1536-1540, Oct. 2019.
	
	\bibitem{Wei2010} S. Wei, D. L. Goeckel, and P. A. Kelly, ``Convergence of the complex envelope of bandlimited OFDM signals,'' {\it IEEE Trans. Info. Theory}, vol. 56, no. 10, pp. 4893-4904, Oct. 2010. 
	
	\bibitem{Harris2008} F. E. Harris, ``Incomplete Bessel, generalized incomplete gamma, or leaky aquifer functions,'' {\em J. Computational and Applied Math.}, vol. 215, no. 1, pp. 260-269, 15 May 2008.
		
	
	\bibitem{Olver2010} F. W. J. Olver, {\em et al.}, {\em NIST Handbook of Mathematical Functions}, Cambridge University Press, New York, NY, 2010. Print companion to {\em NIST Digital Library of Mathematical Functions (DLMF)}: http://dlmf.nist.gov/, Release 1.0.11 of 2016-06-08. 
		
		
\end{thebibliography}
\end{document}